\documentclass[12pt,preprint]{aastex}

\slugcomment{}

\shorttitle{z'--band emission of WASP-12b}
\shortauthors{L\'opez-Morales et al.}

\begin{document}

\title{Day-side z'--band emission and eccentricity of WASP-12b\altaffilmark{1}}

\author{Mercedes L\'opez-Morales\altaffilmark{2,4}, Jeffrey L. Coughlin\altaffilmark{3,5}, David K. Sing\altaffilmark{6}, Adam Burrows\altaffilmark{7}, D\'aniel Apai\altaffilmark{8}, Justin C. Rogers\altaffilmark{4,9}, David S. Spiegel\altaffilmark{7} \& Elisabeth R. Adams\altaffilmark{10}}

\affil{e-mail: mercedes@dtm.ciw.edu}

\altaffiltext{1}{Based on observations collected with the Apache Point
  Observatory 3.5-meter telescope, which is owned and operated by the
  Astrophysical Research Consortium (ARC).}  \altaffiltext{2}{Hubble
  Fellow} \altaffiltext{3}{NSF Graduate Research Fellow}
\altaffiltext{4}{Carnegie Institution of Washington, Department of
  Terrestrial Magnetism, 5241 Broad Branch Rd. NW, Washington D.C.,
  20015, USA} \altaffiltext{5}{Department of Astronomy, New Mexico
  State University, Las Cruces, NM 88003, USA}
\altaffiltext{6}{Astrophysics group, School of Physics, University of
  Exeter, Stocker Road, Exeter, Ex4 4QL, United Kingdom}
\altaffiltext{7}{Princeton University, Department of Astrophysical
  Sciences, Peyton Hall, Princeton, NJ, 08544, USA}
\altaffiltext{8}{Space Telescope Science Institute, 3700 San Martin
  Drive, Baltimore, MD 21218, USA} 
\altaffiltext{9}{Johns Hopkins
  University, Department of Physics and Astronomy, 366 Bloomberg
  Center, 3400 N. Charles Street, Baltimore, MD 21218, USA}
\altaffiltext{10}{Department of Earth, Atmospheric and Planetary Sciences, Massachusetts Institute of Technology, 77 Massachusetts Ave., Cambridge, MA, 02139}

\begin{abstract}

We report the detection of the eclipse of the very-hot Jupiter
\object{WASP-12b} via $z'$-band time-series photometry obtained with
the 3.5-meter ARC telescope at Apache Point Observatory.  We measure a
decrease in flux of 0.082$\pm$0.015\% during the passage of the planet
behind the star. That planetary flux is equally well reproduced by
atmospheric models with and without extra absorbers, and blackbody
models with $f \ge 0.585\pm0.080$. It is therefore necessary to
measure the planet at other wavelengths to further constrain its
atmospheric properties. The eclipse appears centered at phase $\phi
=0.5100^{+0.0072}_{-0.0061}$, consistent with an orbital eccentricity
of $|e \cos \omega| = 0.016^{+0.011}_{-0.009}$ (see note at end of
\S4). If the orbit of the planet is indeed eccentric, the large radius
of WASP-12b can be explained by tidal heating.



\end{abstract}


\section{Introduction} \label{sec:intro}

The transiting Very Hot Jupiter WASP-12b, discovered by
\cite{2009ApJ...693.1920H}, has many notable characteristics. With a
mass of 1.41 $\pm$ 0.10 $M_{Jup}$ and a radius of 1.79 $\pm$ 0.09
$R_{Jup}$, WASP-12b is the planet with the second largest radius
reported to date. It is also the most heavily irradiated planet known,
with an incident stellar flux at the substellar point of over
9$\times10^{9}$ $erg$ $cm^{-2}$ $s^{-1}$. In addition, model fits to
its observed radial velocity and transit light curves suggest that the
orbit of WASP-12b is slightly eccentric. All these attributes make
WASP-12b one of the best targets to test current irradiated atmosphere
and tidal heating models for extrasolar planets.

In irradiated atmosphere model studies WASP-12b is an extreme
case even in the category of highly irradiated gas giants.
Such highly irradiated planets are expected to show thermal inversions
in their upper atmospheric layers \citep{2008ApJ...678.1436B},
although the chemicals responsible for such inversions
remain unknown. TiO and VO molecules, which can act as strong optical
absorbers, have been proposed
\citep{2003ApJ...594.1011H,2008ApJ...678.1419F}, but
\citet{2008A&A...492..585D} claim that the concentration of those
molecules in planetary atmospheres is too low ($< 10^{-3} - 10^{-2}$
times solar). \citet{2009ApJ...699.1487S} argue that TiO needs to
be at least half the solar abundance to cause thermal inversions, and very
high levels of macroscopic mixing are required to keep enough TiO in
the upper atmosphere of planets.  $S_{2}$, $S_{3}$ and HS compounds
have also recently been suggested and then questioned as causes
of the observed thermal inversions
\citep{2009AAS...21430601Z,2009ApJ...701L..20Z}.

In the case of tidal heating, detailed models are now being developed
(e.g. \citealt{2003ApJ...592..555B,2009ApJ...702.1413M,2009arXiv0910.4394I,2009arXiv0910.5928I,2009ApJ...700.1921I})
to explain the inflated radius phenomenon observed in hot
Jupiters, of which WASP-12b, with a radius over 40$\%$ larger than
predicted by standard models, is also an extreme case. All models
assume that the planetary orbits are slightly eccentric, and directly
measuring those eccentricities is key not only to test the model
hypotheses, but also to obtain information about the planets' core
mass and energy dissipation mechanisms (see
e.g. \citealt{2009arXiv0910.4394I}).

Here we present the detection of the eclipse of WASP-12b in the
$z'$-band (0.9 $\mu$m), which gives the first measurement of the
atmospheric emission of this planet, and the first direct estimation
of its orbital eccentricity. Section 2 summarizes the observations and
analysis of the data. In Section 3 we compare the emission of the
planet to models. The results are discussed in Section 4.

\section{Observations and Analyses} \label{sec:obs}

We monitored WASP-12 [RA(J2000)=06:30:32.794, Dec(J2000)=+29:40:20.29,
  V=11.7] during two eclipses, and under photometric conditions, on
February 19 and October 18 2009 UT.  An additional attempt on October
30 2009 UT was lost due to weather. The data were collected with the
SPICam instrument on the ARC's 3.5-meter telescope at Apache Point
Observatory, using a SDSS z$^{\prime}$ filter with an effective
central wavelength of $\sim$0.9$\mu$m. SPICam is a
backside-illuminated SITe TK2048E 2048x2048 pixel CCD with 24 micron
pixels, giving an unbinned plate scale of 0.14 arc seconds per pixel
and a field of view of 4.78 arc minutes square. The detector,
cosmetically excellent and linear through the full A/D converter
range, was binned 2x2, which gives a gain of 3.35 e$^{-}$/ADU, a read
noise of 1.9 DN/pixel, and a 48 second read time.

On February 19 we monitored WASP-12 from 3:00 to 3:28 UT and from 3:54
to 7:10 UT, losing coverage between 3:28 and 3:54 UT when the star
reached a local altitude greater than 85$\degr$, the soft limit of the
telescope at that time. These observations yielded 1.20 hours of
out-of-eclipse and 2.45 hours of in-eclipse coverage, at airmasses
between 1.005--1.412. On October 18 we extended the altitude soft
limit of the telescope to 87$\degr$ and covered the entire eclipse
from 7:05 to 12:45 UT, yielding 2.73 hours of out-of-eclipse and 2.93
hours of in-eclipse coverage, with airmasses between 1.001--1.801. In
both nights we defocused the telescope to a FWHM of $\sim$ 2$\arcsec$
to reduce pixel sensitivity variation effects, and also to allow for
longer integration times, which minimized scintillation noise and
optimized the duty cycle of the observations. Pointing changed by less
than (x,y)=(4,7) pixels in the October 18 dataset, and by less than
(x,y)=(3,12) pixels on February 19, with the images for this second
night suffering a small gradual drift in the $y$ direction throughout
the night. Integration times ranged from 10 to 20 seconds.  Taking
into account Poisson, readout, and scintillation noise, the
photometric precision on WASP-12 and other bright stars in the images
ranged between 0.07--0.15$\%$ per exposure on February 19, and
between 0.05--0.09$\%$ per exposure on October 18.

The field of view of SPICam was centered at RA(J2000)=06:30:25,
Dec(J2000)=+29:42:05 and included WASP-12 and two other isolated stars
at RA(J2000)=06:30:31.8, Dec(J2000)=+29:42:27 and
RA(J2000)=06:30:22.6, Dec(J2000)=+29:44:42, with apparent brightness
and $B-V$ and $J-K$ colors similar to the target.  Each night's
dataset was analyzed independently and the results combined in the
end. The timing information was extracted from the headers of the
images and converted into Heliocentric Julian Days using the IRAF task
{\it setjd}, which has been tested to provide sub-second timing
accuracy.

We corrected each image for bias-level and flatfield effects
using standard IRAF routines. Dark current was
neglegible. DAOPHOT-type aperture photometry was performed in each
frame. We recorded the flux from the target and the comparison stars
over a wide range of apertures and sky background annuli around each
star. We used apertures between 2 and 35 pixels in one-pixel steps
during a first preliminary photometry pass, and 0.05 pixel steps in
the final photometric extraction. To compute the sky background around
each star we used variable width annuli, with inner radii between 35
and 60 pixels sampled in one-pixel steps.

The best aperture and sky annuli combinations were selected by
identifying the most stable (i.e. minimum standard deviation),
differential light curves between each comparison and the target at
phases out-of-eclipse\footnote{We had to iterate on the out-of-eclipse
  phase limits after finding that the eclipse was centered at $\phi$ =
  0.51. Out-of-eclipse was finally defined as phases $\phi$$<$0.45 and
  $\phi$$>$0.57.}. In the February 19 data, the best photometry
results from an aperture radius of 14.7 pixels for both the target and
the comparison stars, and sky annuli with a 52-pixel inner radius and 22-pixel wide. For the October 18 data, 17.9 pixel apertures and
sky annuli with a 45-pixel inner radius and 22-pixel wide produce the
best photometry.

The resultant differential light curves between the target and each
comparison contain systematic trends that can be attributed to either
atmospheric effects, such as airmass, seeing, or sky brightness
variations, or to instrumental effects, such as small changes in the
location of the stars on the detector. Systematics can also be
introduced by instrumental temperature or pressure changes, but those
parameters are not monitored in SPICam. We modeled systematics for
each light curve by fitting linear correlations between each parameter
(airmass, seeing, sky brightness variations, and target position) and
the out-of-eclipse portions of the light curves. All detected trends
are linear and there are no apparent residual color difference
effects. The full light curves are then de-trended using those
correlation fits. In the October 18 dataset, airmass effects are the
dominant systematic, introducing a linear baseline trend with an
amplitude in flux of 0.07$\%$. The February 19 dataset also shows
systematics with seeing and time with a total amplitude of also
0.07$\%$. The systematics on this night were modeled using only the
after-eclipse portion of the light curve, and we consider this dataset
less reliable that the October 18 one. The 18 pre-ingress images
collected between 3:00 and 3:38 UT suffer from a $\sim$50 pixel
position shift with respect to the rest of the images collected that
night, which cannot be modeled using overall out-of-eclipse
systematics. We chose not to use those points in the final analysis.
Correlations with the other parameters listed above are not
significant in any of the two datasets.

Finally, we produce one light curve per night by combining the
de-trended light curves of each comparison. The light curves are
combined applying a weighted average based on the Poisson noise of the
individual light curve points. The result is illustrated in Figure
1. The out-of-eclipse scatter of the combined light curves is 0.11$\%$
for the February 19 data and 0.09$\%$ for the October 18 data.
De-trending significantly improves the systematics, but some
unidentified residual noise sources remain, which we have not been
able to fully model.

\begin{figure}[t]
\epsscale{1.0}
\includegraphics[angle=0, width=3.56in]{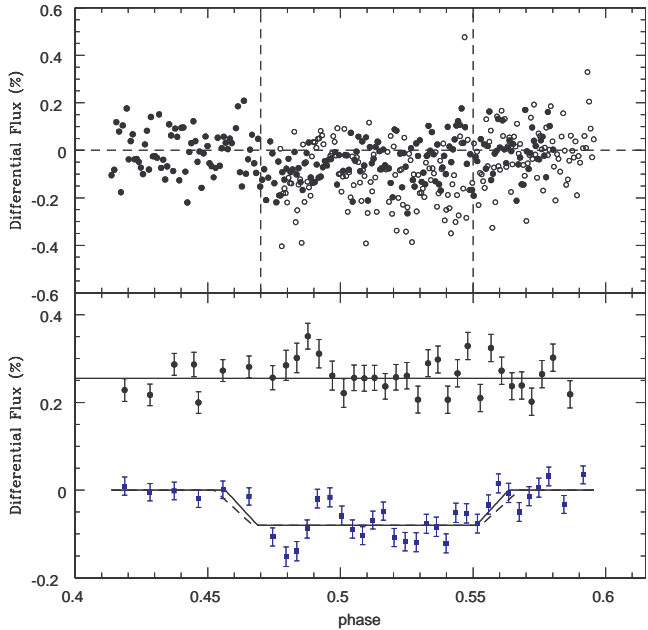}
\caption{{\it Top}: De-trended differential light curves. Open and
  filled dots show, respectively, the Feb 19 and Oct 18 UT 2009
  data. {\it Bottom}: Combined light curves binned by a factor of
  12. Blue squares correspond to WASP-12 and black dots to the
  differential light curve of the two comparison stars. The best fit
  models are shown as solid lines (for $e=0$) and dashed lines (for
  $e=0.057$). Both models produce the same depth and center phase,
  but the $e=0.057$ model lasts 11.52 minutes longer. We attribute the flux jumps between phases 0.475
  and 0.5 to unremoved systematics. Notice that, although the
  systematics appear in both curves, the trends in each curve are not correlated in phase.}
\label{fig:LC}
\end{figure}

\subsection{Eclipse detection and error estimation}

The two-night combined light curve contains 421 points between phases
0.413 and 0.596, based on the \cite{2009ApJ...693.1920H}
ephemerides. To establish the presence of the eclipse and its
parameters, we fit the data to a grid of models generated using the
{\it JASMINE} code, which combines the \citet{2008MNRAS.389.1383K} and
\citet{2002ApJ...580L.171M} algorithms to produce model light curves
in the general case of eccentric orbits. The models do not include limb
darkening and use as input parameters the orbital period, stellar and
planetary radii, argument of the periastron, orbital inclination,
stellar radial velocity amplitude and semi-major axis values derived
by Hebb et al. (2009).  The eccentricity is initially assumed to be
$e$=0, which produces models with a total eclipse duration of 2.808
hours. The best fit model is found by $\chi^{2}$ minimization, with
the depth, the central phase of the eclipse, and the out-of-eclipse
differential flux as free parameters. 

First we fit the individual night light curves to ensure the
eclipse signal is present in each dataset. The February 19 data give
an eclipse depth of 0.100 $\pm$ 0.023$\%$, while the derived eclipse
depth for the October 18 data is 0.068 $\pm$ 0.021$\%$. The central
phases are $\phi$=0.510 for the first eclipse and $\phi$ = 0.508 for
the second. We assume the difference in depth is due to systematics we
have not been able to properly model. The incomplete eclipse from
February 19 might seem more prone to systematics, but our inspection
of both datasets does not reveal stronger trends in that dataset. We
therefore combined the data from both nights, weighting each
light curve based on its out-of-eclipse scatter.

The result of the combined light curve analysis is the detection of an
eclipse with a depth of 0.082 $\pm$ 0.015$\%$ and centered at orbital
phase $\phi$ = 0.51, as shown in Figure 1. The reduced
$\chi^{2}$ of the fit is 0.952. The error in the eclipse depth is
computed using the equation
$\sigma_{depth}^2=\sigma_w^2/N+\sigma_r^2$, where $\sigma_w$ is the
scatter per out-of-eclipse data point and $\sigma_r^2$ describes the
red noise.  The $\sigma_r$ is estimated with the binning technique by
\citet{2006MNRAS.373..231P} to be 1.5$\times10^{-4}$ when binning on
timescales up to the ingress and egress duration of about 20 minutes.

We investigate to what extent the uncertainties in the system's
parameters affect our eclipse depth and central phase results. Varying
the impact parameter, planet-to-star ratio, and scale of the system by
1$\sigma$ of the reported values in \cite{2009ApJ...693.1920H}, the measured
eclipse depth changes only by 0.004$\%$ or 0.27$\sigma_{depth}$, while the
central phase remains unchanged. Our result is therefore largely
independent of the adopted system parameters.

\begin{figure}[b]
\epsscale{1.0}
\includegraphics[angle=0, width=3.39in]{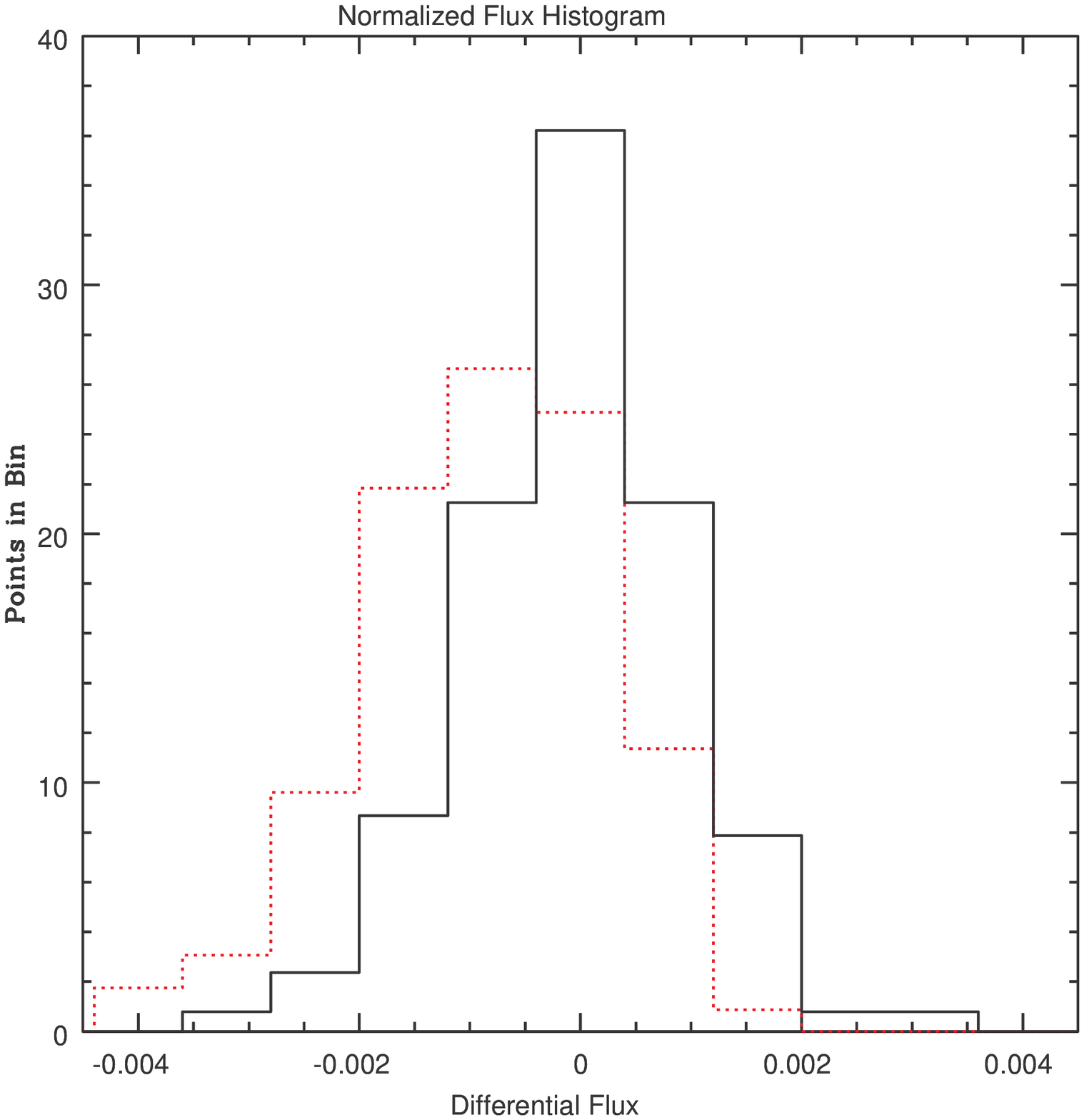}
\caption{Normalized flux histograms of the in-eclipse (dotted red
  line) and out-of-eclipse (solid line) portions of the WASP-12 light
  curve in Figure 1. The bin width is 0.00082 in differential flux,
  coincident with the detected eclipse depth.}
\label{fig:histo}
\end{figure}

\begin{figure}[b]
\epsscale{1.0}
\includegraphics[angle=90, width=3.550in]{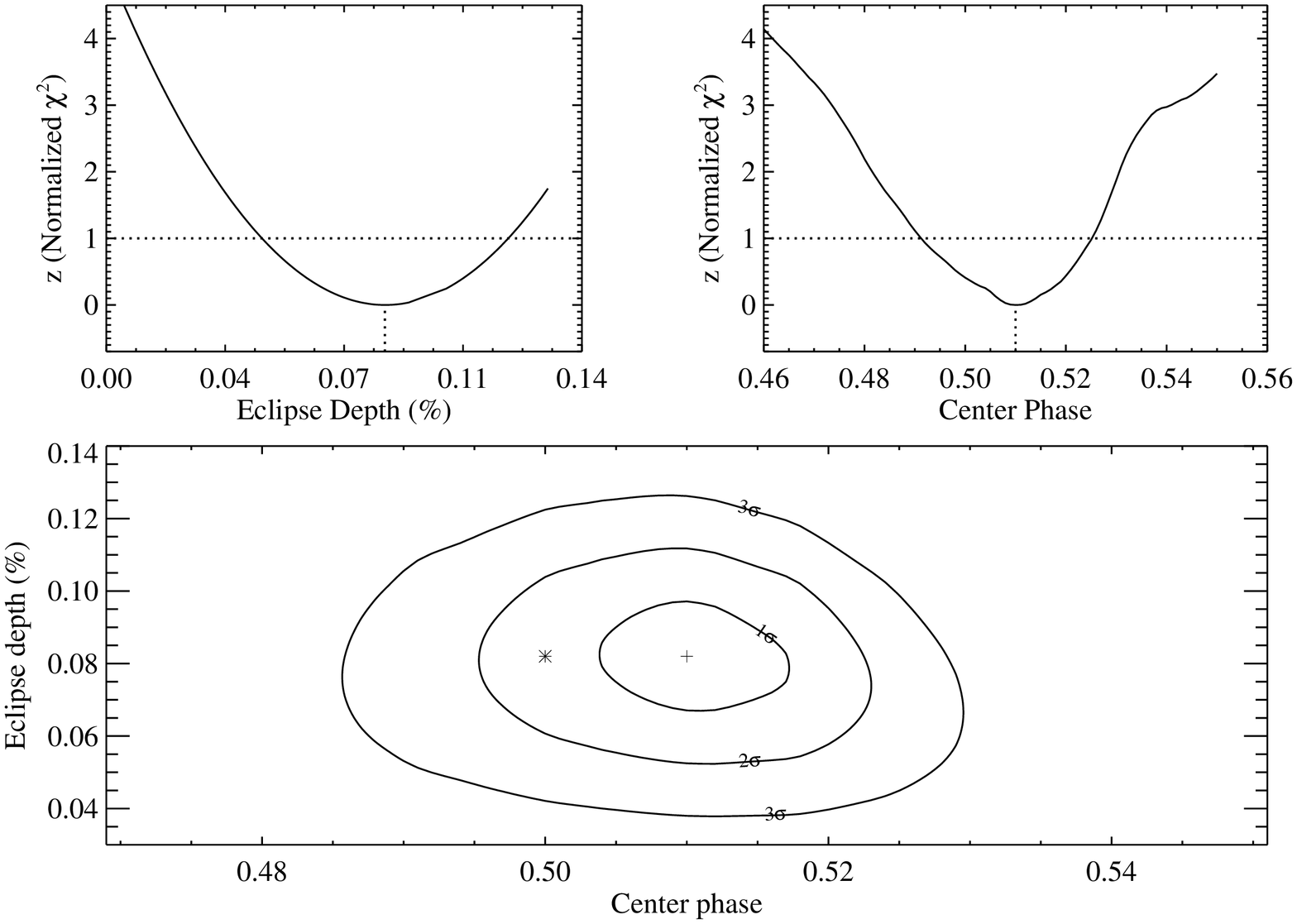}
\caption{{\it Top}: Model eclipse 1--D cuts through the
  normalized--$\chi^{2}$ parameter space for the eclipse depth and
  center phase. $z$=0 gives the best fit and $z$=1 shows the
  1$\sigma$ confidence interval.  {\it Bottom}: Contour plot of the
  eclipse depth versus the center phase, where the 1$\sigma$ depth
  error has been fixed to 0.015$\%$. The best model fit is indicated
  by the cross symbol at $\phi$=0.51 and depth = 0.082$\%$, together
  with 1$\sigma$ to 3$\sigma$ confidence contours. The star symbol at
  $\phi$=0.50 corresponds to a circular orbit.}
\label{fig:contours}
\end{figure}

We performed three other tests to confirm the eclipse detection in a
manner similar to previously reported eclipse results
\citep{2005Natur.434..740D,2009A&A...493L..31S,2009arXiv0910.1257R}.
From the average of the 125 out-of-eclipse light curve data points
versus the 228 in-eclipse points (only points where the planet is
fully eclipsed, adopting $\phi$=0.51 as the central eclipse phase), we
measure an eclipse depth of 0.080 $\pm$ 0.015$\%$.  We further check
the detection by producing histograms of the normalized light curve
flux distribution in the in-eclipse and out-of-eclipse portions of the
light curve. The result, illustrated in Figure 2, shows how the flux
distribution of in-eclipse points is shifted by 0.00082 with respect
to the out-of-eclipse flux distribution, centered at zero. The results
of the last test, where we use a new set eclipse of models with a
duration corresponding to $e=0.057$, fix the out-of-eclipse baseline
to zero, and leave both depth and central phase of the eclipse as a
free parameters, are shown in Figure 3. The two top panels in the
figure show 1-D cuts of each parameter through the $z$ (normalized
$\chi^{2}$), space, where $z$=0 gives the best fit model and $z$=1
defines the 1$\sigma$ confidence interval of the result \citep[see
  definition of $z$ in \S 3
  of][]{2008ApJ...685L..47L,2003ApJS..149...67B}. However, inspection
of the eclipse depth in Figure 1 reveals that the 1$\sigma$ errors
from this method ($\pm$0.036$\%$) are too large. To estimate more
realistic errors for the central phase, we generate contours plots
(bottom panel in Fig. 3), where the 1$\sigma$ eclipse depth error has
been fixed to the 0.015$\%$ value derived above. The resultant center
phase is $\phi = 0.5100^{+0.0072}_{-0.0061}$. We also applied
prayer-bead, bootstrapping, and Markov-Chain Monte Carlo (MCMC) error
analysis techniques \citep{2007A&A...471L..51G, 2008MNRAS.386.1644S}
to estimate the errors in the central phase of the eclipse. The MCMC
analysis contained 2.85$\times10^{6}$ links, and we adopted the
1.5$\times10^{-4}$ photometric error ($\sim$1.6 times the formal
error), given by the \citet{2006MNRAS.373..231P} binning technique.
All these give central phases within $\phi =
0.5100^{+0.0030}_{-0.0036}$. The larger error in the central phase
given by the normalized $\chi^{2}$ method suggests the presence of
correlated noise in the data, which the other methods might not be
correctly accounting for. We therefore adopt these larger errorbars in
our final estimation of the central phase.

\begin{figure}[t]
\epsscale{1.0}
\includegraphics[angle=90, width=3.5in]{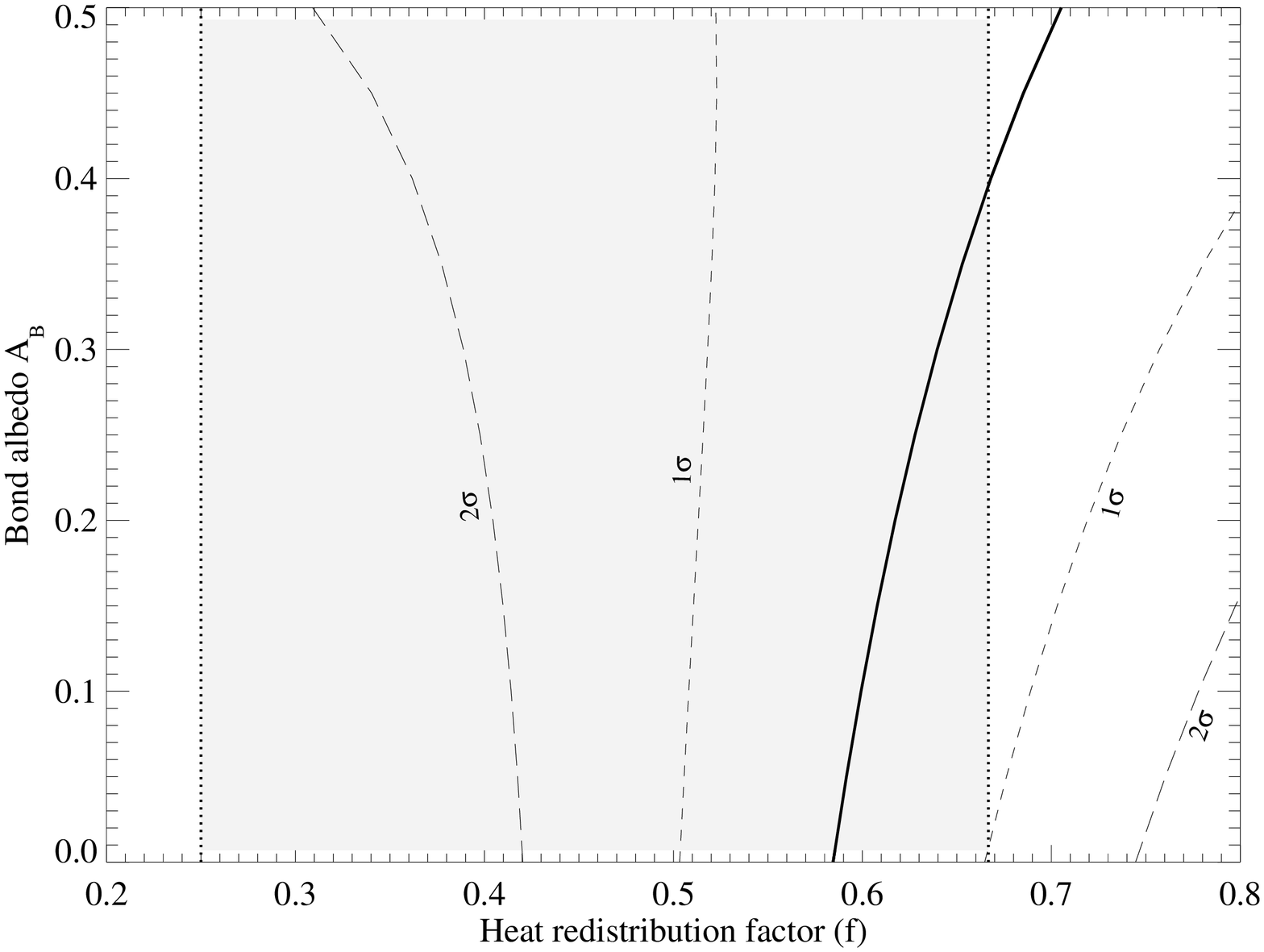}
\caption{Values of $A_{B}$ and $f$ that reproduce the observed
  $z'$-band eclipse depth of WASP-12b, assuming the planet emits as a
  blackbody. The shaded area highlights the region of allowed $f$
  values ($1/4 - 2/3$). The short and long dashed lines delimit,
  respectively, the $1\sigma$ and $2\sigma$ confidence regions of
  the result.}
\label{fig:fvsab}
\end{figure}

\begin{figure}[t]
\epsscale{1.0}
\includegraphics[angle=0, width=3.5in]{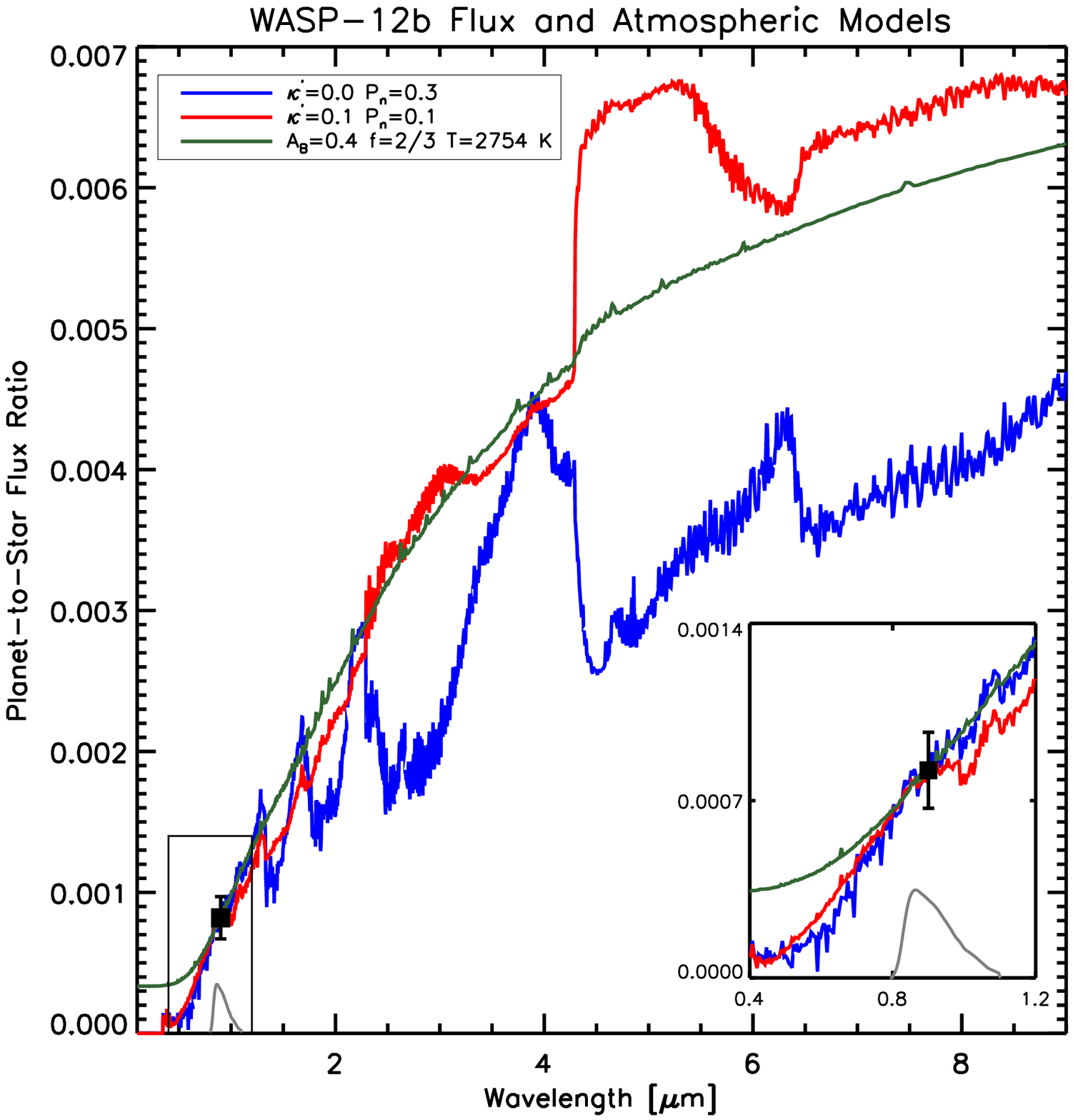}
\caption{Comparison of the eclipse depth (planet-to-star flux ratio)
  to models. The green line shows the $A_{B}$ = 0.4, $f=2/3$ blackbody
  model from Figure 4. The blue and red lines show, respectively, the
  best fit model for an atmosphere with no extra absorber, and with an
  extra absorber of opacity $\kappa'$ between 0.43 and 1.0 $\mu$m. The
  black thin line at the bottom indicates the SPICam plus SDSS
  $z'$-band filter response. See \S3 for more details. }
\label{fig:models}
\end{figure}


\subsection{Eccentricity}

The eccentricity $e$ of WASP-12b was calculated from the measured
central phase shift value using eq. (6) from
\citet{1950ArA.....1...59W},
\begin{equation}
e\cos\omega = \frac{\pi}{P}\frac{(t_{2}-t_{1}-P/2)}{1+\csc^{2}i},
\end{equation}
\noindent where $P$, $i$ and $\omega$ are, respectively, the orbital
period, inclination, and periastron angle of the system, and $t_{2} -
t_{1}$ is the time difference between transit and eclipse. In our
case $t_{2} - t_{1} = 0.51P$. Using the values of $P$, $i$ and
$\omega$ from \citet{2009ApJ...693.1920H}, we derive an $e =
0.057^{+0.040}_{-0.034}$, which agrees with the non-zero eccentricity result
reported by these authors. This eccentricity can be in principle
explained if 1) the system is too young to have already circularized,
2) there are additional bodies in the system pumping the eccentricity
of WASP-12b, or 3) the tidal dissipation factor $Q^{'}_{P}$
\citep{1963MNRAS.126..257G} of WASP-12b is several orders of magnitude
larger than Jupiter's, estimated to be between $6\times10^{4}$ and
$2\times10^{6}$ \citep{1981Icar...47....1Y}.


\section{Comparison with atmospheric models} \label{sec:model}

We compare the observed $z'$-band flux of WASP-12b to simple blackbody
models and to expectedly more realistic radiative-convective models of
irradiated planetary atmospheres in chemical equilibrium, following
the same procedure described in \citet{2009arXiv0910.1257R}. The
results are shown in Figures 4 and 5.

In the simplistic blackbody approximation, a 0.082$\pm$0.015\% deep eclipse
corresponds to a $z'$-band brightness temperature of $T_{z'} \sim
3028^{+99}_{-110} K$, slightly lower than the planet's equilibrium
temperature of $T_{p} \sim 3129 K$ assuming zero Bond albedo
($A_{B}=0$) and no energy reradiation ($f=2/3$) \citep[see][]{2007ApJ...667L.191L}. However, when the thermal and
reflected flux of the planet are included, different combinations of
$A_{B}$ and $f$ can yield the same eclipse depth, as illustrated in
Figure 4. From that figure we can constrain the energy redistribution
factor to $f \ge 0.585 \pm 0.080$, but the albedo is not well
constrained.  Assuming a maximum $A_{B} \le 0.4$, the temperature of
the day-side of WASP-12b is $T_{p} > 2707 K$. 

The more realistic atmospheric models are derived from self-consistent coupled
radiative transfer and chemical equilibrium calculations, based on the
models described in \citet{2000ApJ...538..885S, 2003ApJ...588.1121S},
\citet{2003ApJ...594.1011H} and
\citet{2005ApJ...625L.135B,2006ApJ...650.1140B,2008ApJ...678.1436B}
\citep[see][for details]{2009arXiv0910.1257R}.  We generate models
with and without thermal inversion layers, by adding an unidenfitied
optical absorber between 0.43 and 1.0 $\mu$m, with different level of
opacity $\kappa'$. The opacity of the absorber varies parabolically
with frequency, with a peak value of $\kappa'$ = 0.25 $cm^{2}$
$gr^{-1}$. As Figure 5 shows, models with and without extra
absorbers produce similar fits to the observed $z'$-band flux. The best
model without absorber has a $P_{n} = 0.3$\footnote{$P_{n} = 0$ and
  $P_{n} = 0.5$ correspond, respectively, to $f=2/3$ and $f=1/4$,
  however there is not well defined $P_{n}-f$ relation for
  intermediate values since the physical models account for
  atmospheric parameters (e.g. pressure, opacity) in a way different
  than blackbody models.}. The best model with an extra absorber has a
$P_{n} = 0.1$ and $\kappa_{e}$ = 0.1 $cm^{2}$ $gr^{-1}$. Observations
at other wavelengths are necessary to further constrain the models.

\section{Discussion and Conclusions} \label{sec:sum}

This first detection of the eclipse of WASP-12b agrees with the slight
eccentricity of the planet's orbit found by
\citet{2009ApJ...693.1920H}, and places initial constraints to its
atmospheric characteristics.


 The presence of other bodies in the system
can be tested via radial velocity or transit timing variation
observations, although the current RV curve by
\citet{2009ApJ...693.1920H} shows no evidence of additional planets,
unless they are in very long orbits.

One would expect that if extra absorbers are present in the upper
atmosphere of the planet in gaseous form, they might give rise to thermal
inversion layers.  However, as Figure 5 illustrates, the observed 0.9
$\mu$m eclipse depth can be fit equally well by a model without extra
absorbers. Additional observations at longer wavelengths, specially
longer than $\sim$ 4.0 $\mu$m, will break that model
degeneracy. Observations at wavelengths below $\sim$ 0.6 $\mu$m will
also place better constraints on $A_{B}$.

{\it Note:} While this paper was in the final revision stages, two new
secondary eclipse observations from {\it Spitzer} were reported by
\citet{2010arXiv1003.2763C}. They find secondary eclipse central
phases consistent with a circular orbit. We have carefully reviewed
our analysis and still arrive to a slightly eccentric orbit, although
the result is only significant at the 1.4--1.6$\sigma$ confidence level.
Possible explanations proposed by \citet{2010arXiv1003.2763C} for the
measured eccentricity difference are orbital precession or
wavelength-dependent brightness variations across the surface of the
planet that would shift the center of the eclipse. A similar effect
has been recently suggested by \citet{2010Natur.463..637S} on the
surface of HD189733b.  Further observations are needed to establish
whether this discrepancy is a data artifact or a real effect.

\acknowledgments{M.L.M. acknowledges support from NASA through Hubble
  Fellowship grant HF-01210.01-A/HF-51233.01 awarded by the STScI,
  which is operated by the AURA, Inc. for NASA, under contract
  NAS5-26555. J.L.C acknowledges support from a NSF Graduate Research
  Fellowship.  A.B. is supported in part by NASA grant
  NNX07AG80G. D.A. and J.C.R. are grateful for support from STScI
  Director's Discretionary Research Fund D0101.90131. This work has
  been supported by NSF's grant AST-0908278.}



\clearpage

\end{document}